\documentstyle[amssymb,aps]{revtex}

\begin{document}
\title{Funtional approach to quantum decoherence and the classical final limit II:
the pointer basis and the quantum measurements}
\author{M. Castagnino, R. Laura and R. Id Betan}
\address{Departamenteo de F\'{\i}sica, F.C.E.I.A. Universidad Nacional de Rosario.\\
Instituto de F\'{\i}sica Rosario, CONICET-UNR.\\
Av. Pellegrini 250, 2000 Rosario, Argentina\\
e-mail: castagni@ifir.ifir.edu.ar\\
e-mail: laura@ifir.ifir.edu.ar\\
e-mail: idbetan@ifir.ifir.edu.ar}
\maketitle

\begin{abstract}
A quantum system with discrete and continuos evolution spectrum is studied. A%
{\bf \ final pointer basis} is found, that can be defined in a precised
mathematical way$.$ This result is used to explain the {\bf quantum
measurement} in the system.
\end{abstract}

\section{Introduction}

In the last three years we have devoted some papers to the study of a method
to deal with quantum systems endowed with a continuous evolution spectrum 
\cite{CastagninoL97}\cite{LauraC98PRA}\cite{Castagnino98}\cite{AquilanoCE99}%
\cite{LauraC98PRE}\cite{CastagninoGIBL01}. One of the results of this
research was paper \cite{CastagninoL00} (that can be considered as the part
one of this paper) where decoherence was found in a system with continuous
energy spectrum $0<\omega <\infty $ and a sole discrete eigenvalue $\omega
_{0}<0$. It was demostrated that this system decoheres in a {\it precisely
defined final pointer basis}. The corresponding Wigner functions were
calculated and the classical limit was found. But in this example there was
just one discrete eigenvalue non overlapping with the continuous spectrum,
and therefore it was impossible to use systems of this kind to explain
quantum measurements.

On the contrary in this paper we will solve the problem for a system with a
free Hamiltonian with discrete spectrum with on arbitrary number of
eigenvalues overlapping the continuous spectrum. Again now we will find a
well defined final pointer basis that will be used to explain {\it quantum
measurements} in a very easy way. In paper \cite{CastagninoL00} we have
explained the philosophy of the method, which we do not repeat here for the
sake of conciseness.

The mathematical method that we will use is a generalization of the
perturbation method for systems with continuous spectrum introduced in paper 
\cite{LauraCIB99}. This generalization will be explained in section II. In
section III we will solve the system and find its final pointer basis.
Section IV will be devoted to the use of the system to explain quantum
measurement. We will state our conclusion in section V. An appendix, to show
complementary results, completes the paper.

\section{Perturbative diagonalization of a Liouville operator}

We will generalize section III of paper \cite{LauraCIB99} for the case when
the liouvillian has many discrete eigenvalues. This generalization is quite
straightforward so we will only sketch the main calculations. But let us
begin by reviewing the corresponding Hamiltonians.

We will have \footnote{%
The numbers to the left of the equation corresponds to the equations of
paper \cite{LauraCIB99} and can be used if the reader would like to follow
both papers.} 
\begin{equation}
\begin{array}{ll}
(1_{1}) & \;\;\;\;\;\;H=H^{0}+H^{1}
\end{array}
\,,
\end{equation}
where the free Hamiltonian reads 
\begin{equation}
\begin{array}{ll}
(1_{2}) & \;\;\;\;\;\;H^{0}=\sum\limits_{i=1}^{N}\Omega _{i}|i\rangle
\langle i|+\int\limits_{0}^{\infty }d\omega \,\omega \,|\omega \rangle
\langle \omega |
\end{array}
\,.  \label{2.2}
\end{equation}

The first term of the r.h.s. will play the role of what is usually called
the ``system'' and the second one the role of the ``environment'', $\Omega
_{i}>0$ are the discrete eigenvalues and $0<\omega <\infty $ is the
continuous spectrum. In order to follow the literature as close as possible
we will use the review paper \cite{PazZ99} \footnote{%
Also see bibliography of paper \cite{PazZ99}.} as a base, so we will choose
an interaction similar to $V=\left( \sum\nolimits_{n}\lambda
_{n}\,q_{n}\right) x$, where $q_{n}$ are a set of environment oscillators
and $x$ is the position of a particle (the ``system''). Then in this $V$
there is neither interaction among the environment modes nor the particle
modes. There is only interaction between the environment and the particle.
These properties, translated to our model, give the following interaction
Hamiltonian 
\begin{equation}
\begin{array}{ll}
(1_{3}) & \;\;\;\;\;\;H^{1}=\sum\limits_{i=1}^{N}\,\int\limits_{0}^{\infty
}\,d\omega \,V_{\omega \,i}\,|\omega \rangle \langle
i|+\sum\limits_{i=1}^{N}\,\int\limits_{0}^{\infty }d\omega \,V_{i\,\omega
}\,|i\rangle \langle \omega |
\end{array}
\,,  \label{2.3}
\end{equation}
where $V_{\omega \,i}=\overline{V_{i\,\omega }}$. For simplicity we will
take $V_{\omega \,i}$ real and $V_{\omega \,i}=V_{i\,\omega }$. The basis of
the operators of the system we will use is the following set of vectors
basis 
\begin{eqnarray}
|i\,j) &\equiv &|i\rangle \langle j|\,,\;\;\;\;\;|\omega )\equiv |\omega
\rangle \langle \omega |\,,\;\;\;\;\;|i\,\omega )\equiv |i\rangle \langle
\omega |\,,  \nonumber \\
\,|\omega \,i) &\equiv &|\omega \rangle \langle i|\,,\;\;\;\;\;|\omega
\,\omega ^{\prime })\equiv |\omega \rangle \langle \omega ^{\prime }|\,,
\end{eqnarray}
as in papers \cite{LauraC98PRA}\cite{LauraC98PRE}. The corresponding
co-basis for the state will be 
\begin{equation}
(i\,j|\,,\;\;\;\;\;(\omega |\,,\;\;\;\;\;(i\,\omega |\,,\;\;\;\;\;\,(\omega
\,i|\,,\;\;\;\;\;(\omega \,\omega ^{\prime }|\,
\end{equation}
such that 
\begin{equation}
(i\,j|i^{\prime }\,j^{\prime })=\delta _{i\,i^{\prime }}\delta
_{j\,j^{\prime }}\,,\;\;\;\;\;(\omega |\omega ^{\prime })=\delta (\omega
-\omega ^{\prime })\,,\;\;\;\;\;etc.
\end{equation}

Using these bases the unit superoperator reads 
\begin{eqnarray}
{\Bbb I} &=&\sum\limits_{i\,,\,j}|i\,j)(i\,j|+\int\limits_{0}^{\infty
}d\omega \,|\omega )(\omega |+\sum\limits_{i}\int\limits_{0}^{\infty
}d\omega \,|i\,\omega )(i\,\omega |+  \nonumber \\
&&+\sum\limits_{i}\int\limits_{0}^{\infty }d\omega \,|\omega \,i)(\omega
\,i|+\int\limits_{0}^{\infty }d\omega \int\limits_{0}^{\infty }d\omega
^{\prime }\,|\omega \,\omega ^{\prime })(\omega \,\omega ^{\prime }|\,.
\end{eqnarray}

Also the free liouvillian reads 
\begin{equation}
\begin{array}{ll}
(55) & 
\begin{array}{l}
\;\;\;\;\;\;{\Bbb L}^{0}=\sum\limits_{i\,,\,j}(\Omega _{i}-\Omega
_{j})\,|i\,j)(i\,j|+\sum\limits_{i}\int\limits_{0}^{\infty }d\omega
\,(\Omega _{i}-\omega )|i\,\omega )(i\,\omega |+ \\ 
+\sum\limits_{i}\int\limits_{0}^{\infty }d\omega \,(\omega -\Omega
_{i})|\omega \,i)(\omega \,i|+\int\limits_{0}^{\infty }d\omega
\int\limits_{0}^{\infty }d\omega ^{\prime }\,\,(\omega -\,\omega ^{\prime
})\,|\omega \,\omega ^{\prime })(\omega \,\omega ^{\prime }|\,.
\end{array}
\\ 
& 
\end{array}
\,,
\end{equation}
and the interaction liouvillian is 
\begin{equation}
\begin{array}{ll}
(56) & 
\begin{array}{l}
\;\;\;\;\;\;{\Bbb L}^{1}=\sum\limits_{i\,,\,j}\int\limits_{0}^{\infty
}d\omega \left[ V_{\omega \,i}\,|\omega \,j)-V_{\omega \,j}\,|i\,\omega
)\right] \,(i\,j|+ \\ 
+\sum\limits_{i}\left[ \int\limits_{0}^{\infty }d\omega
\int\limits_{0}^{\infty }d\omega ^{\prime }\,V_{\omega \,i}\,|\omega
\,\omega ^{\prime })-\sum\limits_{j}\int\limits_{0}^{\infty }d\omega
^{\prime }V_{j\,\omega ^{\prime }}|i\,j)\right] \,(i\,\omega ^{\prime }|+ \\ 
+\sum\limits_{i}\left[ \sum\limits_{j}\int\limits_{0}^{\infty }d\omega
\,V_{i\,\omega }\,|i\,j)-\int\limits_{0}^{\infty }d\omega
\int\limits_{0}^{\infty }d\omega ^{\prime }V_{\omega ^{\prime }\,j}|\omega
\,\omega ^{\prime })\right] \,(\omega \,j|+ \\ 
\int\limits_{0}^{\infty }d\omega \int\limits_{0}^{\infty }d\omega ^{\prime
}\sum\limits_{i}\left[ V_{i\,\omega }\,|i\,\omega ^{\prime })-V_{i\,\omega
^{\prime }}\,|\omega \,i)\right] (\omega \,\omega ^{\prime }|+ \\ 
\int\limits_{0}^{\infty }d\omega \sum\limits_{i}\left[ V_{i\,\omega
}\,|i\,\omega )-V_{i\,\omega }\,|\omega \,i)\right] (\omega |
\end{array}
\\ 
& 
\end{array}
\,.
\end{equation}

Now we can repeat all the computations of paper \cite{LauraCIB99} from eq.
(62) to eq. (69) with no changes. The first change appears in eq. (70$_{1}$)
when we must define the proyector on the invariance space of the extended
evolution liouvillian ${\Bbb L}_{ext}^{0}$, that now reads 
\begin{equation}
\begin{array}{ll}
(70_{1}) & \;\;\;\;\;\;{\Bbb P}_{0}=\sum\limits_{i}\,|i\,i)(i\,i|+\int%
\limits_{0}^{\infty }d\omega \,|\omega )(\omega |
\end{array}
\,.  \label{p0}
\end{equation}

Now we can extend ${\Bbb I}$, ${\Bbb L}^{0}$ and ${\Bbb L}^{1}$ to the
complex plane as in paper \cite{LauraCIB99} 
\begin{equation}
\begin{array}{ll}
(59) & 
\begin{array}{l}
\;\;\;\;\;\;{\Bbb I}_{ext}=\sum\limits_{i\,,\,j}\,|i\,j)(i\,j|+\int%
\limits_{0}^{\infty }d\omega \,|\omega )(\omega
|+\sum\limits_{i}\int\limits_{\Gamma }dz^{\prime }\,|i\,z^{\prime
})(i\,z^{\prime }|+ \\ 
+\sum\limits_{i}\int\limits_{\bar{\Gamma}}\,dz\,|z\,i)(z\,i|+\int\limits_{%
\bar{\Gamma}}\,dz\int\limits_{\Gamma }dz^{\prime }\,|z\,z^{\prime
})(z\,z^{\prime }|\,.
\end{array}
\\ 
& 
\end{array}
\,,  \label{iext}
\end{equation}
\begin{equation}
\begin{array}{ll}
(60) & 
\begin{array}{l}
\;\;\;\;\;\;{\Bbb L}_{ext}^{0}=\sum\limits_{i\,,\,j}(\Omega _{i}-\Omega
_{j})\,|i\,j)(i\,j|+\sum\limits_{i}\,\int\limits_{\Gamma }dz^{\prime
}\,(\Omega _{i}-z^{\prime })|i\,z^{\prime })(i\,z^{\prime }|+ \\ 
+\sum\limits_{i}\int\limits_{\bar{\Gamma}}dz\,(z-\Omega
_{i})|z\,i)(z\,i|+\int\limits_{\bar{\Gamma}}dz\int\limits_{\Gamma
}dz^{\prime }\,\,(z-\,z^{\prime })\,|z\,z^{\prime })(z\,z^{\prime }|\,.
\end{array}
\\ 
& 
\end{array}
\,,
\end{equation}
\begin{equation}
\begin{array}{ll}
(61) & 
\begin{array}{l}
\;\;\;\;\;\;{\Bbb L}_{ext}^{1}=\sum\limits_{i\,,\,j}\left[ \int\limits_{\bar{%
\Gamma}}dzV_{z\,i}\,|z\,j)-\int\limits_{\Gamma }\,dz^{\prime }V_{z^{\prime
}\,j}\,|i\,z^{\prime })\right] \,(i\,j|+ \\ 
+\sum\limits_{i}\int\limits_{\Gamma }dz^{\prime }\left[ \int\limits_{\bar{%
\Gamma}}dz\,V_{z\,i}\,|z\,z^{\prime })-\sum\limits_{j}V_{j\,z^{\prime
}}|i\,j)\right] \,(i\,z^{\prime }|+ \\ 
+\sum\limits_{i}\int\limits_{\bar{\Gamma}}dz\left[
\sum\limits_{j}\,V_{i\,z}\,|i\,j)-\int\limits_{\Gamma }dz^{\prime
}V_{z^{\prime }\,j}|z\,z^{\prime })\right] \,(z\,j|+ \\ 
\int\limits_{\bar{\Gamma}}dz\int\limits_{\Gamma }dz^{\prime
}\sum\limits_{i}\left[ V_{i\,z}\,|i\,z^{\prime })-V_{i\,z^{\prime
}}\,|z\,i)\right] (z\,z^{\prime }|+ \\ 
\sum\limits_{i}\int\limits_{\Gamma }dz^{\prime }V_{i\,z^{\prime
}}\,|i\,z^{\prime })-\int\limits_{\Gamma }dzV_{i\,z}\,|z\,i)(z|
\end{array}
\\ 
& 
\end{array}
\,.
\end{equation}
and we can define the ``non-diagonal'' projectors: 
\begin{equation}
{\Bbb P}_{nd}=\sum\limits_{i\neq j}|i\,j)(i\,j|
\end{equation}

Then 
\begin{equation}
\;{\Bbb I}_{ext}={\Bbb P}_{0}+{\Bbb P}_{nd}+{\Bbb P}_{\Gamma }+{\Bbb P}_{%
\bar{\Gamma}}+{\Bbb P}_{\Gamma \,\bar{\Gamma}}  \label{14}
\end{equation}
where the last three projectors correspond to the last three integrals of
eq. (\ref{iext}).

Now we are ready to compute the discrete diagonal spectrum and the
corresponding eigenvectors. Everything will be the same as in paper \cite
{LauraCIB99} e.g. $\lambda _{ii}^{(0)}=\lambda _{ii}^{(1)}=0$ until we reach
eq. (74) that now reads 
\begin{equation}
\begin{array}{ll}
(74) & \;\;\;\;\;\lambda _{ii}^{(2)}|\phi _{ii}^{(0)})=\left[ {\Bbb P}_{0}%
{\Bbb L}_{ext}^{1}{\Bbb Q}_{0}\frac{(-1)}{{\Bbb Q}_{0}{\Bbb L}_{ext}^{0}%
{\Bbb Q}_{0}}{\Bbb Q}_{0}{\Bbb L}_{ext}^{1}{\Bbb P}_{0}\right] |\phi
_{ii}^{0})
\end{array}
\label{lambda2}
\end{equation}
where 
\begin{equation}
{\Bbb Q}_{0}={\Bbb I}_{ext}-{\Bbb P}_{0}={\Bbb P}_{nd}+{\Bbb P}_{\Gamma }+%
{\Bbb P}_{\bar{\Gamma}}+{\Bbb P}_{\Gamma \,\bar{\Gamma}}\,.
\end{equation}

From the definition of ${\Bbb P}_{0}$ (\ref{p0}) we see that only discrete
diagonal terms with vectors ({\it ii}) remain between the square brackets of
eq. (\ref{lambda2}). Then this eigenvalue equation reads 
\begin{equation}
\begin{array}{l}
2\pi i\left[ \sum_{i}\,V_{\Omega _{i}\,i}^{2}\,|i\,i)(i\,i|\phi
_{ii}^{(0)})-V_{\Omega _{i}\,i}^{2}\,|i\,i)(\omega -\Omega _{i})(\phi
_{ii}^{(0)}|\right] = \\ 
\\ 
=\lambda _{ii}^{(2)}\left[ \sum_{i}|i\,i)(i\,i|\phi _{ii}^{(0)})+\int
\,d\omega \,|\omega )(\omega |\phi _{ii}^{(0)})\right] 
\end{array}
\end{equation}
and it has the solutions 
\begin{equation}
\begin{array}{lll}
& \lambda _{d\,ii}^{(2)}=2\pi i\,V_{\Omega _{i}\,\,i}^{2} & \;\;\;\;\;|\phi
_{dii}^{(0)})=|i\,i) \\ 
(77) &  &  \\ 
& \lambda _{\omega }^{(2)}=0 & \;\;\;\;\;|\phi _{\omega
}^{(0)})=\sum_{i}\delta (\omega -\Omega _{i})|i\,i)+|\omega )
\end{array}
\,.  \label{solu}
\end{equation}

Then, as in paper \cite{LauraCIB99}, the degeneration of the discrete
eigenvalues of ${\Bbb L}^{0}$ (namely $\infty \times \infty $) have being
partially removed since now we have just an $\infty $ degeneration for
eigenvalue $\lambda _{\omega }=0$. Going to higher orders it can be shown
that this degeneration always remains. We could have forseen the solution (%
\ref{solu}), since there is no interaction among the discrete modes, there
is only interaction between the discrete and continuous modes, so solution (%
\ref{solu}) must be just the sum of the corresponding solution of paper \cite
{LauraCIB99} (see eq. (77)).

In analogous way we can find the corresponding eigenbras 
\begin{equation}
\begin{array}{lll}
& \lambda _{d\,ii}^{(2)}=2\pi i\,V_{\Omega _{i}\,\,i}^{2} & \;\;\;\;\;(\psi
_{dii}^{(0)}|=(i\,i|-(\omega =\Omega _{i}| \\ 
(84) &  &  \\ 
& \lambda _{\omega }^{(2)}=0 & \;\;\;\;\;(\psi _{\omega }^{(0)}|=(\omega |
\end{array}
\,.  \label{bra}
\end{equation}

The computation of the rest of the spectrum and the eigenbasis follow the
same lines and we obtain 
\begin{eqnarray}
{\Bbb I}_{ext} &=&\int\limits_{0}^{\infty }d\omega \,|\phi _{\omega })(\psi
_{\omega }|+\sum\limits_{i\,,\,j}|\phi _{dij})(\psi
_{dij}|+\sum\limits_{i}\int\limits_{\bar{\Gamma}}du\,|\phi _{u\,i})(\psi
_{u\,i}|  \nonumber \\
&&+\sum\limits_{i}\int\limits_{\Gamma }du^{\prime }|\phi _{i\,u^{\prime
}})(\psi _{i\,u^{\prime }}|+\int\limits_{\bar{\Gamma}}du\int\limits_{\Gamma
}du^{\prime }|\phi _{u\,u^{\prime }})(\psi _{u\,u^{\prime }}|
\end{eqnarray}
\begin{eqnarray}
{\Bbb L}_{ext} &=&\sum\limits_{i\,,\,j}\,\lambda _{d\,ij}\,|\phi
_{dij})(\psi _{dij}|\,+\sum\limits_{i}\int\limits_{\bar{\Gamma}}du\,\lambda
_{u\,i}\,|\phi _{u\,i})(\psi _{u\,i}|  \nonumber \\
&&+\sum\limits_{i}\int\limits_{\Gamma }du^{\prime }\,\lambda _{i\,u^{\prime
}}\,|\phi _{i\,u^{\prime }})(\psi _{i\,u^{\prime }}|+\int\limits_{\bar{\Gamma%
}}du\int\limits_{\Gamma }du^{\prime }\,\lambda _{u\,u^{\prime }}\,|\phi
_{u\,u^{\prime }})(\psi _{u\,u^{\prime }}|  \label{ldiago}
\end{eqnarray}
where up to second order for the eigenvalues: 
\begin{eqnarray}
\,\lambda _{d\,ij} &=&\Omega _{i}-\Omega _{j}+i\pi (V_{\Omega
_{i}\,i}^{2}-V_{\Omega _{j}\,j}^{2})+  \nonumber \\
&&+\int\limits_{0}^{\infty }d\omega \,\left[ -V_{\omega \,i}^{2}\,{\cal P}%
\frac{1}{\omega -\Omega _{i}}+V_{\omega \,j}^{2}{\cal P}\frac{1}{\omega
-\Omega _{j}}\right] 
\end{eqnarray}
and for $i=j$ the corresponding eigenvectors are 
\begin{eqnarray}
|\phi _{dii}) &=&|i\,i)+\int\limits_{0}^{\infty }d\omega \,\frac{V_{\omega
\,i}}{\Omega _{i}-\omega }|\omega \,i)+\int\limits_{0}^{\infty }d\omega \,%
\frac{V_{\omega \,i}}{\Omega _{i}-\omega }|i\,\omega ) \\
(\psi _{dii}| &=&(i\,i|-(\omega =\Omega _{i}|  \label{24}
\end{eqnarray}
these are the only new results (see appendix for the corresponding
calculations). The rest is a simple straight generalization. Precisely 
\begin{eqnarray}
\lambda _{u\,i} &=&u-\Omega _{i} \\
|\phi _{u\,i}) &=&|u\,i)+\frac{V_{u\,i}}{u-\Omega _{i}}|i\,i)-\int\limits_{%
\Gamma }du^{\prime }\frac{V_{u^{\prime }\,i}}{u^{\prime }-\Omega _{i}}%
|u\,u^{\prime }) \\
(\psi _{u\,i}| &=&(u\,i|+\frac{V_{u\,i}}{u-\Omega _{i}}\left[
(i\,i|-(u|\right] 
\end{eqnarray}
\begin{eqnarray}
\lambda _{i\,u^{\prime }} &=&\Omega _{i}-u^{\prime }+i\pi V_{\Omega
_{i}\,i}^{2}-\int\limits_{0}^{\infty }d\omega \,V_{\omega \,\,i}^{2}{\cal P}%
\frac{1}{\omega -\Omega _{i}} \\
|\phi _{i\,u^{\prime }}) &=&|i\,u^{\prime })+\frac{V_{u^{\prime }\,i}}{%
u^{\prime }-\Omega _{i}}|i\,i)-\int\limits_{\bar{\Gamma}}du\frac{V_{u\,i}}{%
u-\Omega _{i}}|u\,u^{\prime }) \\
(\psi _{i\,u^{\prime }}| &=&(i\,u^{\prime }|+\frac{V_{u^{\prime }\,i}}{%
u^{\prime }-\Omega _{i}}\left[ (i\,i|-(u^{\prime }|\right] -\int\limits_{%
\bar{\Gamma}}du\frac{V_{u\,i}}{u-\Omega _{i}}(u\,u^{\prime }|
\end{eqnarray}
\begin{eqnarray}
\lambda _{u\,u^{\prime }} &=&u-u^{\prime } \\
|\phi _{u\,u^{\prime }}) &=&|u\,u^{\prime })+\sum_{i}\left[ \frac{V_{u\,i}}{%
u-\Omega _{i}}|i\,u^{\prime })+\frac{V_{u^{\prime }\,i}}{u^{\prime }-\Omega
_{i}}|u\,i)\right]  \\
(\psi _{u\,u^{\prime }}| &=&(u\,u^{\prime }|+\sum_{i}\left[ \frac{V_{u\,i}}{%
u-\Omega _{i}}(i\,u^{\prime }|+\frac{V_{u^{\prime }\,i}}{u^{\prime }-\Omega
_{i}}(u\,i|\right] 
\end{eqnarray}

So now that the generalization is completed we will see the physical
consequences of these, somehow unfamiliar, equations in the next section.

\section{Decoherence and the final pointer basis}

Let us first study the decoherence in the energy and then the one in all the
observable of a CSCO that contains the Hamiltonian.

\subsection{Decoherence in the energy}

From paper \cite{CastagninoL00} we know that a generalized state (let say at 
$t=0$) reads 
\begin{eqnarray}
(\rho (0)| &=&\int d\omega \,\rho _{\omega }\,(\psi \,_{\omega
}|+\sum_{i\,j}\,\rho _{d\,i\,j}\,(\psi \,_{d\,i\,j}|+  \nonumber \\
&&\sum_{i}\left[ \int_{\bar{\Gamma}}d\omega \,\rho _{\omega \,i}\,(\psi
\,_{\omega \,i}|+\int_{\Gamma }d\omega ^{\prime }\,\rho _{i\,\omega ^{\prime
}}\,(\psi \,_{i\,\omega ^{\prime }}|\right] +  \nonumber \\
&&+\int_{\bar{\Gamma}}d\omega \int_{\Gamma }d\omega ^{\prime }\,\rho
_{\omega \,\omega ^{\prime }}(\psi \,_{\omega \,\omega ^{\prime }}|
\end{eqnarray}
where the $\rho $ satisfies the conditions stated in the quoted paper $\rho
_{\omega }=\overline{\rho _{\omega }}\geqslant 0\,$; $\rho _{d\,i\,j}=%
\overline{\rho _{d\,j\,i}}\,$; $\rho _{d\,i\,i}\geqslant 0\,$; $\rho
_{\omega \,i}=\overline{\rho _{i\,\omega }}\,\,$; $\rho _{\omega \,\omega
^{\prime }}=\overline{\rho _{\omega ^{\prime }\,\omega }}$ \footnote{%
In paper \cite{LauraC98PRA} for the sake of simplicity was postulated that $%
\rho _{\omega }$ was a regular function. Now we will change this asumption
and allow to $\rho _{\omega }$ to contain Diracs deltas as $\delta (\omega
-\Omega _{i})$.}, and 
\[
\int_{0}^{\infty }d\omega \,\rho _{\omega }+\sum_{i}\rho _{d\,i\,i}=1\,.
\]

Using the diagonalized liouvillian of eq. (\ref{ldiago}) we can write the
same state at $t$ (up to second perturbation order in the eigenvalues) as: 
\begin{eqnarray}
(\rho (t)| &=&\int d\omega \,\rho _{\omega }\,(\omega |+  \nonumber \\
&&+\sum_{i\,j}\,\rho _{d\,i\,j}\,(\psi \,_{d\,i\,j}|\,\exp \left\{ t\left[
-\pi \left( V_{d\,i}^{2}-V_{d\,j}^{2}\right) +i\left( \Omega _{i}-\Omega
_{j}\right) +\right. \right.  \nonumber \\
&&\left. \left. i\int d\omega \left( -V_{\omega \,i}^{2}\,{\cal P}\frac{1}{%
\omega -\Omega _{i}}+V_{\omega \,j}^{2}\,{\cal P}\frac{1}{\omega -\Omega _{j}%
}\right) \right] \right\} +  \nonumber \\
&&+\sum_{i}\int_{\bar{\Gamma}}\,du\,\rho _{u\,i}\,(\psi
_{u\,i}|\,e^{i(u-\Omega _{i})\,t}+  \nonumber \\
&&+\sum_{i}\int_{\Gamma }\,du^{\prime }\,\rho _{i\,u^{\prime }}\,(\psi
_{i\,u^{\prime }}|\,e^{i(\Omega _{i}-u^{\prime })\,t}\,e^{-\pi V_{\Omega
_{i}\,i}^{2}\,t}\,\exp \left( -i\,\int_{0}^{\infty }d\omega \,\frac{%
V_{\omega \,i}^{2}}{\omega -\Omega _{i}}\,t\right) +  \nonumber \\
&&+\int_{\bar{\Gamma}}du\int_{\Gamma }\,du^{\prime }\,e^{i(u-u^{\prime
})\,t}\,(\psi _{u\,u^{\prime }}|
\end{eqnarray}

Taking into account the Riemann-Lebesgue theorem and the dumping factors we
have: 
\begin{equation}
\lim_{t\rightarrow \infty }(\rho (t)|=\int d\omega \,\rho _{\omega
}\,(\omega |+\,\sum_{i}\,\rho _{d\,ii}\,(\psi \,_{d\,i\,i}|\,=\rho _{*}
\label{limitro}
\end{equation}
begin $\rho _{*}$ the equilibrium final state.

So, as in paper \cite{CastagninoL00}, the state $\rho (t)$ decoheres into a
diagonal state of the continuous part of the spectrum. Thus, we can ask
ourselves where the terms are corresponding to the discrete part of the
spectrum of the free liouvillian 
\begin{equation}
(\rho (0)|=\sum_{i\,j}\,\rho _{d\,i\,j}^{0}\,(i\,j|\,
\end{equation}
when $t\rightarrow \infty $. They are dissolved in the continuous spectrum
but they are still there as we will see. While the off-diagonal terms $i\neq
j$ have disappeared as it should, dumped by their evolution factor, the
diagonal terms $i=j$ are in $\rho _{\omega }$. To understand what is going
on let us begin by a simple case. Let us suppose that $t=0$ there are only
discrete terms, and let us write the diagonal terms of $(\rho (0)|$ in the
eigenbasis of ${\Bbb L}$%
\begin{equation}
\sum_{i}\,\rho _{\,ii}^{0}\,(i\,i|\,=\int d\omega \,\rho _{\omega }\,(\omega
|\,+\sum_{i}\,\rho _{d\,i\,i}\left[ (i\,i|-(\omega =\Omega _{i}|\right] 
\label{rocero}
\end{equation}
where we have used eq. (\ref{24}) to compute $(\psi _{d\,i\,i}|$. Then we
necessarily have 
\begin{eqnarray}
\rho _{d\,i\,i} &=&\rho _{i\,i}  \nonumber \\
\rho _{\omega } &=&\sum_{i}\rho _{\,i\,i}^{0}\,\delta (\omega -\Omega _{i})
\label{3.7}
\end{eqnarray}
as it can easily be verified. Thus from eq. (\ref{limitro}) we have 
\begin{equation}
\lim_{t\rightarrow \infty }(\rho (t)|=\int d\omega \sum_{i}\rho
_{\,i\,i}^{0}\,\delta (\omega -\Omega _{i})(\omega |\,=\sum_{i}\rho
_{\,i\,i}^{0}\,(\omega =\Omega _{i}|=\rho _{*}  \label{3.8}
\end{equation}

So we have found the{\bf \ final pointer basis} $\left\{ (\omega =\Omega
_{i}|\right\} ${\bf \ }in this case i.e. the pointer basis for the discrete
part of the spectrum.

If the initial conditions would be completely general, with discrete and
continuous diagonal components, etc., i.e. 
\begin{equation}
(\rho (0)|=\sum_{i\,j}\,\rho _{\,i\,j}^{0}\,(i\,j|\,+\int d\omega \,\rho
_{\omega }^{0}\,(\omega |\,+...  \label{3.8p}
\end{equation}
(where $(ij|$ and $(\omega |$ are now considered as eigenvectors of ${\Bbb L}%
_{0}$ and the dots symbolize other of diagonal terms) we would have in the
basis of ${\Bbb L}$%
\begin{equation}
(\rho (0)|=\sum_{i}\,\rho _{d\,i\,i}\,\left[ (i\,i|-(\omega =\Omega
_{i}|\,\right] +\int d\omega \,\rho _{\omega }\,(\omega |\,+...
\end{equation}
so 
\begin{eqnarray}
\rho _{d\,i\,i} &=&\rho _{\,i\,i}^{0} \\
\rho _{\omega } &=&\sum_{i}\rho _{\,i\,i}^{0}\delta (\omega -\Omega
_{i})+\rho _{\omega }^{0}
\end{eqnarray}

So in this general case we have 
\begin{equation}
\lim_{t\rightarrow \infty }(\rho (t)|=\sum_{i}\rho _{\,i\,i}^{0}\,(\omega
=\Omega _{i}|+\int d\omega \sum_{i}\rho _{\,\omega }^{0}\,(\omega |\,=\rho
_{*}  \label{3.8bis}
\end{equation}
since all the off-diagonal terms (the dots) disappear as explained so the 
{\bf whole final pointer basis} is $\left\{ (\omega |\,,(\omega =\Omega
_{i}|\right\} $.

As an illustration we can compute the time evolution of the diagonal terms
(discrete and continuous) for the simple initial condition (\ref{rocero}).
The evolution equation is 
\begin{equation}
(\rho (t)|=(\rho (0)|\,e^{\,i\,{\Bbb L\,}t}\,
\end{equation}
and all its elements must be projected under the diagonal projector 
\begin{equation}
{\Bbb P}=\int d\omega \,|\phi _{\omega })(\psi _{\omega }|+\sum_{i}|\phi
_{d\,i\,i})(\psi _{d\,i\,i}|
\end{equation}

We will add a subindex $d$ to the projected diagonal states. Then if 
\[
(\rho (0)|_{d}=\sum_{i}\rho _{d\,i\,i}(\psi _{d\,i\,i}|+\int d\omega \,\rho
_{\omega }\,(\psi _{\omega }|\,
\]
it turns out that 
\begin{eqnarray}
(\rho (t)|_{d} &=&\sum_{i}\rho _{d\,i\,i}(\psi _{d\,i\,i}|\,\exp (i\,\lambda
_{d\,i\,i}\,t)+\int d\omega \,\rho _{\omega }\,(\psi _{\omega }|=  \nonumber
\\
&=&\,\sum_{i}\rho _{\,i\,i}^{0}\,(i\,i|\,\exp (-2\pi V_{\Omega
_{i}\,i}^{2}\,t)+\,\sum_{i}\rho _{\,i\,i}^{0}\,\left[ 1-\exp (-2\pi
V_{\Omega _{i}\,i}^{2}\,t)\right] \,(\omega =\Omega _{i}|
\end{eqnarray}
where we have used eqs. (\ref{3.7}), (\ref{solu}) and (\ref{bra}). Now we
can clearly see how the factor $\exp (-2\pi V_{\Omega \,i}^{2}\,t)$ produces
the decay of the discrete states $(i\,i|$ while the factors $\left[ 1-\exp
(-2\pi V_{\Omega \,i}^{2}\,t)\right] $ make grow the states in the final
discrete pointer basis: $\{(\omega =\Omega _{i}|\}$ as it should be.

\subsection{Decoherence in the other dynamical variables of the CSCO}

If, as in paper \cite{CastagninoL00} we could have other observable $%
\{O_{m}\}$ in our CSCO (where $m=1,2,..,M$ and we consider that the spectra
of these $O_{m}$ observable are discrete for simplicity). Then the initial
state (that generalizes the one in eq. (\ref{3.8p})) would be 
\begin{equation}
(\rho (0)|=\sum_{i\,j\,m\,m^{\prime }}\rho _{i\,j\,m\,m^{\prime
}}(i\,j\,m\,m^{\prime }|\,+\sum_{m\,m^{\prime }}\int_{0}^{\infty }d\omega
\,\rho _{\omega \,m\,m^{\prime }}\,(\omega \,m\,m^{\prime }|+...
\end{equation}
where the dots symbolize terms in the other vector of the cobasis. Then we
would end as in (\ref{3.8}) with 
\begin{equation}
(\rho _{*}|=\sum_{i\,m\,m^{\prime }}\rho _{i\,i\,m\,m^{\prime }}(\omega
=\Omega _{i}\,m\,m^{\prime }|\,+\sum_{m\,m^{\prime }}\int_{0}^{\infty
}d\omega \,\rho _{\omega \,m\,m^{\prime }}\,(\omega \,m\,m^{\prime }|+...
\end{equation}

Now, following, step-by-step, section IIB of paper \cite{CastagninoL00} we
can diagonalize the matrices $\rho _{i\,i\,m\,m^{\prime }}$ and $\,\rho
_{\omega \,m\,m^{\prime }}$ to obtain 
\begin{equation}
\rho _{*}=\sum_{i\,r}\rho _{i\,i\,r\,r}(\omega =\Omega
_{i}\,r\,r|\,+\sum_{r}\int_{0}^{\infty }d\omega \,\rho _{\omega
\,r\,r}\,(\omega \,r\,r|+...
\end{equation}
where $r=1,2,...,M$ and where $\{(\omega \,r\,r|,(\omega =\Omega
_{i}\,r\,r|\}$ is the final pointer basis that now corresponds to the final
pointer CSCO $\{H,P_{1},...,P_{M}\}$ where the $P_{i}$ have only discrete
spectra by the assumptions made at the beginning of the section, and as in
eq. (22) of paper \cite{CastagninoL00} they read 
\begin{equation}
P_{j}=\sum_{i}P_{i\,r}^{(j)}\,(\omega =\Omega _{i}\,r\,r|+\int_{0}^{\infty
}d\omega \,P_{\omega \,r}^{(j)}(\omega \,r\,r|
\end{equation}

For the sake of simplicity, as in paper \cite{CastagninoL00}, we can make 
\begin{equation}
P_{i\,r}^{(j)}=P_{\omega
\,r}^{(j)}=r^{(j)}=(r_{1}^{(j)},r_{2}^{(j)},...,r_{M}^{(j)})
\end{equation}
in such a way that the eigenvalues $r^{(j)}$ label the eigenstates of $P_{j}$%
. In this way the final pointer CSCO is defined.

\subsection{Wigner function}

Always following paper \cite{CastagninoL00} it can be proved that the Wigner
function corresponding to $\rho _{*}$ reads 
\begin{equation}
\rho _{*}^{W}(q,p)=\sum_{i\,r}\rho _{i\,i\,r\,r}\,\rho
_{i\,r}^{W}(q,p)+\sum_{r}\int_{0}^{\infty }d\omega \,\rho _{\omega
\,r\,r}\,\rho _{\omega \,r}^{W}(q,p)
\end{equation}
where 
\begin{equation}
\rho _{i\,r}^{W}(q,p)=C\,\delta \left[ H^{W}(q,p)-\Omega _{i}\right]
\,\delta \left[ P_{1}^{W}(q,p)-r_{1}\right] ...\delta \left[
P_{M}^{W}(q,p)-r_{M}\right]  \label{3.25a}
\end{equation}
\begin{equation}
\rho _{\omega \,r}^{W}(q,p)=C\,\delta \left[ H^{W}(q,p)-\omega \right]
\,\delta \left[ P_{1}^{W}(q,p)-r_{1}\right] ...\delta \left[
P_{M}^{W}(q,p)-r_{M}\right]
\end{equation}

Therefore we see that the classical analogue of the final pointers basis, $%
\rho _{\omega \,r}^{W}$ and $\rho _{i\,r}^{W}$, are densities which are
peaked along classical trayectories defined by the constant of motion $%
(\omega \,r_{1}...\,r_{M}|$ and $(\Omega _{i}\,r_{1}...\,r_{M}|$
respectively. Moreover they are $\geqslant 0$.

So everything that was said in paper \cite{CastagninoL00} for just one
discrete eigenvalue $\omega _{0}$ can be now repeated by the finite set of
discrete eigenvalues $\Omega _{1}...\Omega _{N}$, e.g. the
histories-decoherence version of the theory (\cite{CastagninoL00} Appendix
C) can be repeated word by word.

\section{Quantum Measurement Theory}

In this section, as an application we will sketch the elements and results
of quantum measurement theory using the results of the previous sections.

As explained in review paper \cite{PazZ99}, that we take as a guide, let us
consider a system with eigenvectors $|s_{i}\rangle $ and a measurement
apparatus with eigenvectors $|A_{i}\rangle $, where now the index $i$
represents the set of indices $i,r$ or better $i,r_{1},...,r_{M}$ of the
previous section. Now we know that the measurement evolution brings a
generic state of the system and the apparatus $|\psi _{0}\rangle =|\psi
\rangle |A_{0}\rangle $ to a new one $|\psi _{t}\rangle $, precisely 
\begin{equation}
|\psi _{0}\rangle =|\psi \rangle |A_{0}\rangle =\left(
\sum_{i}a_{i}|s_{i}\rangle \right) |A_{0}\rangle \rightarrow
\sum_{i}a_{i}|s_{i}\rangle |A_{i}\rangle =|\psi _{t}\rangle 
\end{equation}

So this premeasurement process just correlates the state of the system with
those of the apparatus. The corresponding density matrix evolves under this
process as 
\begin{equation}
\rho _{0}=|\psi _{0}\rangle \langle \psi _{0}|\rightarrow |\psi _{t}\rangle
\langle \psi _{t}|=\rho _{t}
\end{equation}

So the final state is as pure as the initial one. This is precisely the
quantum measurement problem: we must explain how this pure state $\rho _{t}$
evolves to a diagonal density matrix, in such a way to allow
classical-boolean measurements. The problem is readily solved if we call $%
|s_{i}\rangle |A_{i}\rangle =|i\rangle $ and we consider that really the
hamiltonian at the whole ``system'' (i.e. system, plus apparatus) is the
first term of the r.h.s. of eq. (\ref{2.2}) and to which we have added an
``environment'' with an hamiltonian corresponding to the second term.
Essentially the environment is the universe, so it is natural to endow this
system with a continuous spectrum, since the universe contains at least the
electromagnetic and the gravitational fields (both with continuous spectra).
To this $H_{0}$ we can add the $H^{1}$ of eq. (\ref{2.3}) that would be the
simplest coupling possible.

Then, as proved in eq. (\ref{3.8}) the matrix 
\begin{equation}
\rho _{0}=|\psi _{0}\rangle \langle \psi _{0}|=\sum_{i\,j}\bar{a}%
_{i}\,a_{j}|i\rangle \langle j|=\sum_{i\,j}\bar{a}_{i}\,a_{j}(i\,j|
\end{equation}
evolves to the final diagonal density matrix 
\begin{equation}
\rho _{*}=\sum |a_{i}|^{2}\,(\omega =\Omega _{i}|
\end{equation}
in the discrete final pointer basis $\{(\omega =\Omega _{i}|\}$. Moreover
the corresponding Wigner functions of the vectors of this pointer basis are
the ones displayed in eq. (\ref{3.25a}) which capture their classical
nature. In this way the measurement quantum process and its classical output
is completely explained.

\section{Conclusion}

As systems with continuous spectrum are very frequent in the universe if is
clear that they cannot be excluded from the environment. Moreover an
environment endowed with a continuous spectrum is very useful since if
allows the use of Riemann-Lebesgue theorem. Nevertheless a big problem was
that in such a system only the continuous diagonal states are stationary and
therefore the only candidates for final equilibrium states are there. What
is the ultimate fate of the states of the discrete spectrum is an intriguing
question that deseaved an answer. The answer is given in this paper. Such
states dissolve themselves in the continuous diagonal but they remain as
Dirac's deltas originating the pointer basis $\{(\omega =\Omega _{i}|\}$.
With this answer in hand it is easy to foresee the evolution of the system
and to introduce a simple explanation for the quantum measurement process.

\section{Acknowledgment}

This paper was partially supported by grants PID-0150 of CONICET (Argentina
Research Council) and EX-198 of Buenos Aires University. One of us (M.C.)
would like to acknowledge the hospitality of Imperial College (London) where
this paper was begun.

\appendix 

\section{Pertubation}

The only really new characters in this paper with respect to those of paper 
\cite{LauraCIB99} are the discrete off-diagonal elements with $i\neq j$.
Then let us compute their eigenvalues and eigenvectors with the pertubation
method of this paper.

At zero order we have

\begin{equation}
\begin{array}{ll}
& \lambda _{i\,j}^{(0)}=\Omega _{i}-\Omega _{j} \\ 
(85)\;\;\;\;\; & |\phi _{d\,i\,j}^{(0)})=|i\,j) \\ 
& (\psi _{d\,i\,j}^{(0)}|=(i\,j|
\end{array}
\end{equation}

At first order we must solve 
\begin{equation}
\begin{array}{ll}
(63)\;\;\;\;\; & (\lambda ^{(0)}-{\Bbb L}_{ext}^{0})|\phi ^{(1)})=({\Bbb L}%
_{ext}^{1}-\lambda ^{(1)})|\phi ^{(0)})
\end{array}
\end{equation}
that in this case reads 
\begin{equation}
(\Omega _{i}-\Omega _{j}-{\Bbb L}_{ext}^{0})|\phi ^{(1)})=({\Bbb L}%
_{ext}^{1}-\lambda ^{(1)})\,|i\,j)
\end{equation}
which multiplied by $(i^{\prime }\,j^{\prime }|$ gives 
\begin{equation}
(\Omega _{i}-\Omega _{j}-\Omega _{i^{\prime }}+\Omega _{j^{\prime
}})(i^{\prime }\,j^{\prime }|\phi ^{(1)})=-\lambda ^{(1)}\,\delta
_{i\,i^{\prime }}\,\delta _{j\,j^{\prime }}
\end{equation}
because from (\ref{2.3}) $(i^{\prime }\,j^{\prime }|{\Bbb L}%
_{ext}^{1}|i\,j)=0$ since there are no term $|i^{\prime }\,j^{\prime
})(i\,j| $ in ${\Bbb L}_{ext}^{1}$. So if $i=i^{\prime }$, $j=j^{\prime }$
(even if $i\neq j$) we have $\lambda ^{(1)}=0$.

Now from eq. (\ref{2.3}) we have 
\begin{equation}
(\Omega _{i}-\Omega _{j}-{\Bbb L}_{ext}^{0})|\phi ^{(1)})=\int_{0}^{\infty
}\,d\omega \,V_{\omega \,i}\,|\omega \,j)-\int_{0}^{\infty }\,d\omega
\,V_{\omega \,j}\,|i\,\omega )
\end{equation}
so 
\begin{equation}
|\phi _{ij}^{(1)})=\int_{\Gamma }\,\frac{du\,V_{u\,i}}{\Omega _{i}-u}%
\,|u\,j)-\int_{\bar{\Gamma}}\,\frac{du^{\prime }\,V_{u^{\prime }\,j}}{%
u^{\prime }-\Omega _{j}}\,|i\,u^{\prime })\,.
\end{equation}

We now can go to the second order where eq. (64) of paper \cite{LauraCIB99}
gives 
\begin{equation}
(\lambda ^{(0)}-{\Bbb L}_{ext}^{0})|\phi ^{(2)})=({\Bbb L}_{ext}^{1}-\lambda
^{(1)})\,|\phi ^{(1)})-\lambda ^{(2)}|\phi ^{(0)})
\end{equation}
which premultiplied by $(i^{\prime }\,j^{\prime }|$ reads 
\begin{equation}
(\Omega _{i}-\Omega _{j}-\Omega _{i^{\prime }}+\Omega _{j^{\prime
}})(i^{\prime }\,j^{\prime }|\phi ^{(2)})=(i^{\prime }\,j^{\prime }|{\Bbb L}%
_{ext}^{1}|\phi ^{(1)})-\lambda ^{(2)}\,\delta _{i\,i^{\prime }}\,\delta
_{j\,j^{\prime }}\,.
\end{equation}

\begin{eqnarray}
(i^{\prime }\,j^{\prime }|{\Bbb L}_{ext}^{1}|\phi _{i^{\prime \prime
}\,j^{\prime \prime }}^{(1)}) &=&\sum_{i\,j}\int_{\Gamma }du^{\prime
}\,V_{j\,u^{\prime }}\delta _{i\,i^{\prime }}\,\delta _{j\,j^{\prime
}}\delta _{i\,i^{\prime \prime }}\int_{\bar{\Gamma}}\frac{du^{\prime \prime
}\,V_{u^{\prime \prime }\,j^{\prime \prime }}}{u^{\prime \prime }-\Omega
_{j^{\prime \prime }}}\delta (u^{\prime }-u^{\prime \prime })+ \\
&&+\sum_{i\,j}\int_{\bar{\Gamma}}du^{\prime }\,V_{i\,u^{\prime }}\delta
_{i\,i^{\prime }}\,\delta _{j\,j^{\prime }}\delta _{j\,j^{\prime \prime
}}\int_{\Gamma }\frac{du^{\prime \prime }\,V_{u^{\prime \prime }\,i^{\prime
\prime }}}{\Omega _{i^{\prime \prime }}-u^{\prime \prime }}\delta (u^{\prime
\prime }-u^{\prime }) \\
&=&\int_{\Gamma }du^{\prime }\frac{V_{j^{\prime }\,u^{\prime
}}\,V_{u^{\prime }\,j^{\prime \prime }}}{u^{\prime }-\Omega _{j^{\prime
\prime }}}+\int_{\bar{\Gamma}}du^{\prime }\frac{V_{i\,u^{\prime
}}\,V_{u^{\prime }\,i^{\prime \prime }}}{\Omega _{i^{\prime \prime
}}-u^{\prime }}\,.
\end{eqnarray}

Now making $i=i^{\prime }$, $j=j^{\prime }$ we have 
\begin{equation}
\lambda ^{(2)}=\int_{\Gamma }du\frac{V_{u\,j}^{2}}{u-\Omega _{j}}+\int_{\bar{%
\Gamma}}du\frac{V_{u\,i}^{2}}{\Omega _{i}-u}\,.
\end{equation}

But 
\begin{equation}
\int_{\Gamma }du\frac{V_{u\,j}^{2}}{u-\Omega _{j}}=\int_{0}^{\infty }d\omega 
\frac{V_{\omega \,j}^{2}}{\omega -i0-\Omega _{j}}=i\,\pi \,V_{\Omega
_{j}\,j}^{2}+\int_{0}^{\infty }d\omega \,V_{\omega \,j}^{2}{\em P}\left( 
\frac{1}{\omega -\Omega _{j}}\right)
\end{equation}
\begin{equation}
\int_{\bar{\Gamma}}du\frac{V_{u\,i}^{2}}{\Omega _{i}-u}=i\,\pi \,V_{\Omega
_{i}\,i}^{2}-\int_{0}^{\infty }d\omega \,V_{\omega \,i}^{2}{\em P}\left( 
\frac{1}{\omega -\Omega _{i}}\right)
\end{equation}
(cfr. \cite{LauraCIB99} eqs. (89) and (91)). Then 
\begin{equation}
\lambda ^{(2)}=i\,\pi \left[ V_{\Omega _{j}\,j}^{2}+\,V_{\Omega
_{i}\,i}^{2}\right] +\int_{0}^{\infty }d\omega \,\left[ V_{\omega \,j}^{2}%
{\em P}\left( \frac{1}{\omega -\Omega _{j}}\right) -V_{\omega \,i}^{2}{\em P}%
\left( \frac{1}{\omega -\Omega _{i}}\right) \right]
\end{equation}
and the computation is finished.

\end{document}